\documentclass[aps,prl,epsfig,graphics,floatfix,mathbbm,twocolumn]{revtex4}

\usepackage{amsmath,amsfonts,amssymb,graphics,graphicx,epsfig,color,times,bbm}

\begin{document}
\bibliographystyle{apsrev}

\newcommand{\R}{\mathbbm{R}}
\newcommand{\rr}{\mathbbm{R}}
\newcommand{\E}{{\cal E}}
\newcommand{\cc}{{\cal{C}}}
\newcommand{\ii}{\mathbbm{1}}

\newcommand{\1}{\mathbbm{1}}

\newcommand{\tr}{{\rm tr}}
\newcommand{\gr}[1]{\boldsymbol{#1}}
\newcommand{\be}{\begin{equation}}
\newcommand{\ee}{\end{equation}}
\newcommand{\bea}{\begin{eqnarray}}
\newcommand{\eea}{\end{eqnarray}}
\newcommand{\ket}[1]{|#1\rangle}
\newcommand{\bra}[1]{\langle#1|}
\newcommand{\avr}[1]{\langle#1\rangle}
\newcommand{\D}{{\cal D}}
\newcommand{\eq}[1]{Eq.~(\ref{#1})}
\newcommand{\ineq}[1]{Ineq.~(\ref{#1})}
\newcommand{\sirsection}[1]{\section{\large \sf \textbf{#1}}}
\newcommand{\sirsubsection}[1]{\subsection{\normalsize \sf \textbf{#1}}}
\newcommand{\ack}{\subsection*{\normalsize \sf \textbf{Acknowledgements}}}
\newcommand{\front}[5]{\title{\sf \textbf{\Large #1}}
\author{#2 \vspace*{.4cm}\\
\footnotesize #3}
\date{\footnotesize \sf \begin{quote}
\hspace*{.2cm}#4 \end{quote} #5} \maketitle}
\newcommand{\eg}{\emph{e.g.}~}

\newcommand{\proofend}{\hfill\fbox\\\medskip }


\newtheorem{theorem}{Theorem}
\newtheorem{proposition}{Proposition}

\newtheorem{lemma}{Proposition}

\newtheorem{definition}{Definition}
\newtheorem{corollary}{Corollary}

\newcommand{\proof}[1]{{\bf Proof.} #1 $\proofend$}

\title{Area laws in quantum systems: mutual information and correlations}

\author{Michael M. Wolf$^1$, Frank Verstraete$^2$, Matthew B. Hastings$^3$, J. Ignacio Cirac$^1$}
\affiliation{$^1$ Max-Planck-Institut f\"ur Quantenoptik,
Hans-Kopfermann-Str.1, 85748 Garching, Germany.\\$^2$ Fakult\"{a}t
f\"{u}r Physik, Universit\"{a}t
Wien, Boltzmanngasse 5, A-1090 Wien, Austria.\\
$^3$ Center for Non-linear Studies and Theoretical Division, Los
Alamos National Laboratory, Los Alamos, New Mexico 87545, USA}
\date{\today}

\begin{abstract}

The holographic principle states that on a fundamental level the
information content of a region should depend on its surface area
rather than on its volume.  In this paper we show that this
phenomenon not only emerges in the search for new Planck-scale
laws but also in lattice models of classical and quantum physics:
the information contained in part of a system in thermal
equilibrium obeys an area law. While the maximal information per
unit area depends classically only on the number of degrees of
freedom, it may diverge as the inverse temperature in quantum
systems. It is shown that an area law is generally implied by a
finite correlation length when  measured in terms of the mutual
information.

\end{abstract}

\maketitle



\begin{figure}[t]
\resizebox{5.5cm}{!} {\includegraphics{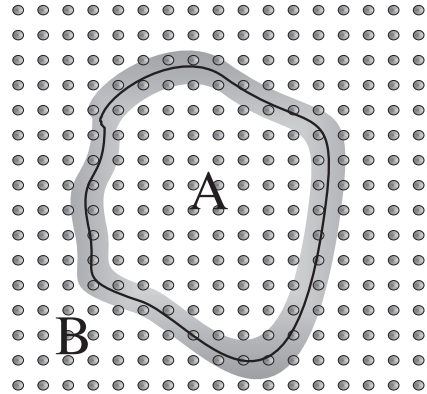}} \caption{ We
are interested in the mutual information (or entanglement) between
the two regions $A$ and $B$. Heuristically, if there is a
correlation length $\xi$ then sites in $A$ and $B$ that are
separated by more than $\xi$ (the shaded stripe) should not
contribute to the information or entanglement between $A$ and $B$.
The mutual information (or entanglement) is thus bounded by the
number of sites at the boundary.} \label{Fig_arealaw}
\end{figure}

Correlations are information of one system about another.
The study of correlations
in equilibrium lattice models comes in two flavors. The more
traditional approach is the investigation of the decay of
two-point correlations with the distance. A lot of knowledge has
been acquired in Condensed Matter Physics in this direction and is
now being used and developed further in the study of entanglement
in Quantum Information Theory \cite{Os02,Os02b,Ve04}. The second
approach (see Fig.\ref{Fig_arealaw}) asks how correlations between
a connected region and its environment scale with the size of that
region. This question has recently been addressed for a variety of
quantum systems at zero temperature
\cite{Pl05,Bo04,Ca05,Vi03,Ji04,Ca04,Fer1,Fer2,Bo86,Sr93} where all
correlations are due to entanglement which in turn is then
measured by the entropy.

The original interest in this topic \cite{Bo86,Sr93,Ho85,Ca94}
came from the insight that the entropy of black holes scales with
the area of the surfaces at the event horizon---we say that an
area law holds, in this case with a maximal information content of
one bit per Planck area. Remarkably, a similar entropy scaling is
observed in non-critical quantum lattice systems while critical
systems are known to allow for small (logarithmic) deviations
\cite{Ca05,Vi03,Ji04,Ca04,Fer1,Fer2}. Both is in sharp contrast to
the behavior of the majority of states in Hilbert space which
exhibit a volume scaling rather than an area law. These insights
fruitfully guided recent constructions of powerful classes of
ansatz states which are tailored to cover the relevant aspects of
strongly correlated quantum many-body systems \cite{Vid05,Ve04a}.

A heuristic explanation of the area law in non-critical systems
comes from the existence of a characteristic length scale, the
correlation length, on which two-point correlations decay
(Fig.\ref{Fig_arealaw}). Intuitively this apparent localization of
correlations should imply an area law, an argument which can,
however, not easily be made rigorous
A firm connection between the decay of
correlations and the area law is thus still lacking as well as is
a proof and extension of the latter beyond zero temperature. In
the present work we address both problems by resorting to a
concept of Quantum Information Theory---the mutual information.
The motivation for this quantity is that (i) it coincides with the
entanglement entropy at zero temperature; (ii) it measures the
total amount of information of one system about another without
'overlooking' hidden correlations; (iii) the area law can be
rigourously proven at any finite temperature; (iv) the heuristic
picture relating decay of correlations and area law can be made
rigorous in the form of a one-way implication.
 Moreover, we will prove that an area law is fulfilled by
all mixed projected entangled pair states (PEPS), discuss the
behavior  of the mutual information for certain classes of 1D
systems in more detail, and show that a strict 1D-area law implies
that the state has an exact representation as a finitely
correlated state.

We begin by fixing some notation. We consider systems on lattices
$\Lambda\subseteq\mathbb{Z}^\D$ in $\D$ spatial dimensions which
are sufficiently homogeneous (e.g., translational invariant). Each
site of the lattice corresponds to a classical or quantum spin
with configuration space $\mathbb{Z}_d$ or Hilbert space
$\mathbb{C}^d$ respectively. Given a probability distribution
$\rho$ on $\Lambda$ and marginals $\rho_A,\rho_B$ corresponding to
disjoint sets $A,B\subseteq\Lambda$, the mutual information
between these regions is defined by
\be
\label{eq:MIdef}I(A:B)=
H(\rho_A)+H(\rho_B)-H(\rho_{AB}),
\ee
where $H(\rho)=-\sum_{x}\rho(x)\log\rho(x)$ is the Shannon
entropy. In the quantum case the $\rho$'s become density operators
(and their partial traces) and $H$ has to be replaced by the von
Neumann entropy $S(\rho)=-\tr[\rho\log\rho]$. The mutual
information has a well defined operational meaning as the total
amount of correlations between two systems \cite{Gr05}. It
quantifies the information about $B$ which can be obtained from
$A$ and vice versa. Elementary properties of the mutual
information are positivity, that it vanishes iff the system
factorizes, and it is non-increasing under discarding parts of the
system \cite{NielsenChuang}. We will occasionally write $S_A$
meaning $S(\rho_A)$.\vspace*{5pt}

\emph{Area laws for classical and quantum systems:} Let us start
considering classical Gibbs distributions of finite range
interactions. All such distributions are Markov fields, i.e., if
$x_A,x_C,x_B$ are configurations of three regions where $C$
separates $A$ from $B$ such that no interaction directly connects
$A$ with $B$, then $\rho(x_A|x_C,x_B)=\rho(x_A|x_C)$ holds for all
conditional probabilities [with $\rho(x|y):=\rho(x,y)/\rho(y)$].
Let us denote by $\partial A$, $\partial B$ the sets of sites in
$A,B$ which are connected to the exterior by an interaction.
Exploiting the Markov property together with the fact that we can
express the mutual information in terms of a conditional entropy
$H(A|B)=H(A)-I(A:B)$ then leads to an area law
\be
I(A:B)=I(\partial A: \partial B)\leq H(\partial A) \leq
|\partial A| \log d,
\label{eq:boundaryC}
\ee
where the first inequality follows from positivity of the
classical conditional information. Equation (\ref{eq:boundaryC})
shows that correlations in classical thermal states are localized
at the boundary. In particular if we take $B$ the complement of
$A$, then we obtain that the mutual information scales as the
boundary area of the considered region and the maximal information
per unit area is determined by the number of microscopic degrees
of freedom.

For quantum systems less information can be inferred from the
boundary and the Markov property does no longer hold in general.
Remarkably enough, for the case of the mutual information between
a region $A$ and its complement $B$ we can also derive an area law
for finite temperatures. In order to show that, we consider again
a finite range Hamiltonian $H=H_A+H_\partial + H_B$, where $H_A$,
$H_B$ are all interaction terms within the two regions and
$H_\partial $ collects all those crossing the boundary. The
thermal state $\rho_{AB}$ corresponding to the inverse temperature
$\beta$ minimizes the free energy $F(\rho)=\tr[H\rho]-\frac1\beta
S(\rho)$. In particular, $F(\rho_{AB})\leq F(\rho_A\otimes\rho_B)$
from which we obtain
\be \label{eq:QthermalArea} I(A:B)\leq \beta\;
\tr\big[H_\partial(\rho_A\otimes\rho_B-\rho_{AB})\big]
\ee
since $H_A,H_B$ have the same expectation values in both cases. As
the r.h.s. of Eq.(\ref{eq:QthermalArea}) depends solely on the
boundary we obtain again an area law scaling similar to that in
Eq.(\ref{eq:boundaryC}). For example, if we just have two--site
interactions we obtain $I(A:B)\le 2 \beta ||h|| |\partial A|$,
where $||h|||$ is the maximal eigenvalue of all two--site
Hamiltonians across the boundary, i.e., the strength of the
interaction. Note that the scale at which the area law becomes
apparent is now determined by the inverse temperature $\beta$. In
fact, it is known that at zero temperature the boundary area
scaling of the mutual information, which then becomes  $I(A:B)=2
S(A)$, breaks down for certain critical systems
\cite{Ca05,Vi03,Ji04,Ca04,Fer1,Fer2}. Eq.(\ref{eq:QthermalArea})
shows that all the logarithmic corrections appearing in these
models disappear at any finite temperature.

By comparing the area laws (\ref{eq:boundaryC}) and
(\ref{eq:QthermalArea}) we notice that quantum states may have
higher mutual information than classical ones as the information
per unit area is no longer bounded by the number of degrees of
freedom. In fact, our results imply that if a system violates
inequality (\ref{eq:boundaryC}), then it must have a quantum
character. Note that
Eqs.(\ref{eq:boundaryC},\ref{eq:QthermalArea}) directly generalize
the findings of \cite{Cr06} for systems of harmonic oscillators.

Let us now turn to an important class of quantum states which goes
beyond Gibbs states, namely projected entangled pair states (PEPS)
\cite{Ve04a}. These states bear their name from projecting
`virtual spins', obtained from assigning entangled pairs
$|\Phi\rangle=\sum_{i=1}^D|ii\rangle$ to the edges of a lattice,
onto physical sites corresponding to the vertices. A natural
generalization of this concept to mixed states is to use
completely positive maps for the mapping from the virtual to the
physical level \cite{Ve04b}. Since every such map can be purified,
these mixed PEPS can be interpreted as pure PEPS with an
additional physical system which gets traced out in the end.
For all these states
one can now easily see that the mutual information between a block
$A$ and its complement $B$ satisfies a boundary area law
\be
I(A:B)\leq 2 |\partial A|\log
D,
\label{eq:PEPSarea}
\ee
since it is upper bounded by the mutual information, i.e., twice
the block entropy, of the purified state which is in turn bounded
by the number of  bonds cut. An interesting class of mixed PEPS
are Gibbs states of Hamiltonians of commuting finite range
interactions (see appendix). Note that these are not necessarily
classical systems, as a simultaneous diagonalization need not
preserve the local structure of the interaction. The Kitaev model
\cite{Ki03} on the square lattice, the cluster state \cite{Ra01}
Hamiltonian and all stabilizer Hamiltonians fall in this class.
Moreover, Gibbs states of arbitrary local Hamiltonians are
approximately representable as mixed PEPS \cite{mbh}.\vspace*{5pt}

\emph{Mutual information and correlations: } We will now discuss
the correlations measured in terms of the mutual information
between separate regions. Traditionally, correlations are measured
by connected correlation functions $\cc(M_A,M_B):=\langle
M_A\otimes M_B\rangle -\langle M_A\rangle \langle M_B \rangle$ of
observables $M_A$, $M_B$. In fact, these two concepts can be
related by expressing the mutual information as a relative entropy
$S(\rho_{AB}|\rho_A\otimes\rho_B)=I(A:B)$,  using the norm bound
$S(\rho|\sigma)\geq\frac12||\rho-\sigma||_1^2$ \cite{Petz} and the
 inequality $||X||_1\geq\tr[XY]/||Y||$. In this way we obtain
\be
I(A:B)\geq
\frac{\cc(M_A,M_B)^2}{2||M_A||^2\||M_B||^2}\label{eq:twopoint}.
\ee
Hence, if $I(A:B)$ decays for instance exponentially in the
distance between $A$ and $B$ then so will $\cc$. One of the
advantages of the mutual information is,  that there cannot be
correlations `overlooked', whereas $\cc$ might be arbitrarily
small while the state is still highly correlated---a fact
exploited in quantum data hiding and quantum expanders\cite{Ha04}.

\begin{figure}[t]
\center \resizebox{7cm}{!}{\includegraphics{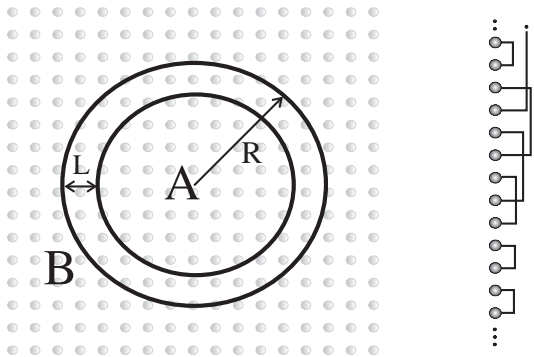}}
\caption{Left: We consider regions $A$ and $B$ separated by a
spherical shell of thickness $L\ll R$; Right: Simple 1D model for
a state which is formed by singlet pairs (indicated by lines
joining them) whose length follows a given probability
distribution.} \label{Fig_correlations}
\end{figure}

In the following we will relate the correlation length as defined
by the mutual information with the area law mentioned previously.
To this end consider a spherical shell $C$ of outer radius $R$ and
thickness $L\ll R$ which separates the inner region $A$ from the
exterior $B$ (see Fig. \ref{Fig_correlations}). We denote the
mutual information between $A$ and $B$ by $I_L(R)$ and define
$\xi_{M}$ as the minimal length $L$ such that $I_L(R)<I_0(R)/2$
for all $R$, i.e., a correlation length measured by the mutual
information. Note that $\xi_M$ can be infinite (e.g., for critical
systems) and that it takes into account the decay of all possible
correlations. Using the subadditivity property of the entropy we
obtain the general inequality $I(A:BC)\leq I(A:B)+2S_C$ which
leads to
\be I_0\leq I_{\xi_M} + 2 S_C \leq 4 |\partial A| \xi_{M}.
\label{eq:expt2area} \ee
Here the first inequality implies the second one by inserting
$I_{\xi_M}\leq\frac12 I_0$ and the fact that $S(C)\le
\xi_M\;|\partial A|$ . So, indeed, we get an area law for the
mutual information solely from the existence of the length scale
$\xi_M$, which expresses the common sense explanation of Fig.
\ref{Fig_arealaw}. This area law is also valid for zero
temperature and when violated immediately implies an infinite
correlation length $\xi_M$. The converse is, however, not true
since there are critical lattice systems which obey an area law
\cite{PEPSarea,Cr06}. Surprisingly, an area law can even hold
under algebraically decaying two-point correlations
\cite{PEPSarea,Cr06}.\vspace*{5pt}

\emph{Examples in one dimension: } We will now investigate the
decay of correlations in terms of the mutual information for
certain simple cases. We will show that in all of them $\xi_M$ is
directly connected to the standard correlation length. We will
consider infinite lattices in 1 spatial dimension (see Fig.
\ref{Fig_correlations}).

We start out by considering an important class of states, the
so--called finitely correlated states (FCS) \cite{FNW92}, which
naturally appear in several lattice systems in 1D. They can be
viewed as 1D PEPS (or matrix product states). Every FCS is most
easily characterized by a completely positive, trace preserving
map (a channel) $T:{\cal B(H}_1)\rightarrow {\cal B(H}_1\otimes
{\cal H}_2)$ with ${\cal H}_1,{\cal H}_2$ Hilbert spaces of
dimension $D, d$ respectively. Define further $\E(x)=\tr_2[T(x)]$
and assume the generic condition that $\E$ has only one eigenvalue
of magnitude one. The second largest eigenvalue, $\eta$, is
related to the standard correlation length through
$\xi\sim-1/\ln\eta$. In order to estimate $\xi_M$ we exploit the
fact that $\rho_{AB}$ factorizes exponentially with increasing
separation $L$, i.e., $||\rho_{AB}-\rho_A\otimes \rho_B||_1={\cal
O}(e^{-L/\xi})$ (see appendix). Moreover, $T$ can be locally
purified thereby increasing the size of ${\cal H}_2$ by a factor
of $d D^2$. Denoting the additional purifying systems by $A'$ and
$B'$ respectively, we obtain on the one hand $I(A:B)\leq
I(AA':BB')=S(\rho_{AA'}\otimes\rho_{BB'})-S(\rho_{AA'BB'})$. On
the other hand we can apply Fannes' inequality\cite{Fannes},
$|S(\rho)-S(\sigma)|\leq \Delta\log (\delta-1) +
H(\Delta,1-\Delta)$, where $\Delta=\frac12||\rho-\sigma||_1$ and
$\delta$ is the dimension of the supporting Hilbert space, to
$I(AA':BB')$. Due to the purification we deal with finite
dimensional systems ($\delta=D^2$) so that putting things together
leads to
\be I_L(R)\leq \log(D) {\cal O}\Big(L\;e^{-L/\xi}\Big).
\label{eq:FCSexp} \ee
Since $I_L(R)$ increases (decreases) with $R$ ($L$), and is lower
bounded by correlation function (\ref{eq:twopoint}) this
inequality immediately implies that $\xi_M$ is finite and directly
related to  $\xi$.

The  case considered above includes several interesting situations
of systems in 1D with finite--range interactions:
frustration--free Hamiltonians at $T=0$, all classical Gibbs
states, and all quantum ones for commuting Hamiltonians. In all
cases, the area law is fulfilled following the results given in
the previous sections. However, it is known that for certain
critical local Hamiltonians the area law is violated at $T=0$. In
order to analyze how this behavior may emerge, we will considered
a simple toy model in 1D  for which $I_L(R)$ can be exactly
determined.

Let us consider a spin $\frac12$ system formed of singlets (see
Fig. \ref{Fig_correlations}). The state is such that from any
given site, $i$, the probability of having a singlet with another
site, $j$, is a function $f(|i-j|)$. The mutual information
between two regions is equal to the number of singlets that
connect those regions, and thus it can be easily determined (if we
take a large region, so that we can average this number). If we
take $f(x)\propto e^{-x/\xi}$ we have that: (i) all (averaged)
correlation functions decay exponentially with the distance and
that $\xi$ gives the correlation length; (ii) $I_L(R)$ decays
exponentially with $L$ and that $\xi_M\sim\xi$; (iii) an area law
is fulfilled. If we take $f(x)\propto 1/(x^2+a^2)$ we obtain that:
(i) the correlation functions decay as power laws with the
distance; (ii) $I_L(R)\sim \log(2R-L)$ and thus $\xi_M$ is
infinite; (iii) the area law is violated. Thus, for this specific
model we see how the violation of the area law naturally implies
an infinite correlation length.

For zero temperature there is another simple connection between
the area law and the decay of $I_L(R)$ as a function of the
separation $L$. If for a pure state the entropy of a block of
length $L$ goes to a constant $K$ as $S_L=K-f(L)$ with
$f(L)\rightarrow 0$ for increasing $L$, then $I_L(R)\to f(L)$ as
$R\to\infty$ for sufficiently large $L$. If the block entropy
diverges instead, then $I_L(R)\to \infty$ for every finite
separation.\vspace*{5pt}

\emph{Saturation of mutual information implies FCS:} For
one-dimensional systems the area law just means a saturation of
the mutual information. Let us finally gain some first insight
into the structure of states having this property. So consider a
general (mixed) 1D translational invariant state and denote the
mutual information between a block of length $L$ and the rest of
the system by $I(L)$ and similarly its entropy by $S(L)$. The
latter can be shown to be a concave function
\be S(L)\geq \big(S(L-1)+S(L+1)\big)/2 \label{eq:ConcaveS}, \ee
which is nothing but the strong subadditivity inequality applied
to a region of length $L-1$ surrounded by two single sites.
Eq.(\ref{eq:ConcaveS}) has strong implications on the behavior of
$I(L)$. Assume that the system is a finite ring of length $N$,
then
\bea I(L)-I(L-1)\ \label{eq:Idiff}
&=&[S(L)-S(L-1)]\\&&-[S(N-L+1)-S(N-L)]\nonumber \eea
is a difference between two slopes of the entropy function. Due to
concavity of $S(L)$, $I(L)$ is increasing as long as $L<N/2$.
Moreover, if from some length scale on the mutual information
exactly saturates, i.e., $I(L-1)=I(L)$ then all slopes between $L$
and $N-L$ have to be equal so that strong subadditivity in
Eq.(\ref{eq:ConcaveS}) holds with equality. States with this
property are, however, nicely characterized \cite{Ha04b} and known
to be quantum Markov chains. That is, there exists a channel
$\tilde{T}:{\cal B\big(H}_2^{\otimes(L-1)}\big)\rightarrow {\cal
B\big(H}_2^{\otimes L}\big)$ such that \be ({\rm id}\otimes
\tilde{T})(\rho_{L-1})=\rho_{L}\label{eq:QMarkov},\ee where
$\rho_L$ is the reduced density operator of $L$ sites and
successive applications of $\tilde{T}$ to the last $L-1$ sites
generates larger and larger parts of the chain. For infinite
systems these states form a subset of the FCS where now
$D=d^{(L-1)}$, i.e., the scale at which saturation sets in
determines the ancillary dimension needed to represent the state.
\vspace*{5pt}

\emph{Acknowledgements: } Portions of this work were done at the
ESI-Workshop on Lieb-Robinson Bounds. MBH was supported by US DOE
DE-AC52-06NA25396. We acknowledge financial support by the
European projects SCALA and CONQUEST and by the DFG
(Forschungsgruppe 635 and Munich Center for Advanced Photonics
(MAP)).

\section{Appendix}

In this appendix we show that (i) every Gibbs state of a local
quantum Hamiltonian with mutually commuting interactions is a
mixed projected entangled pair state (PEPS) with small bond
dimension, and (ii) finitely correlated states factorize
exponentially, i.e., exhibit an exponential split property.

\subsection{PEPS representation of thermal stabilizer states}

PEPS \cite{Ve04a} bear their name from projecting `virtual spins',
obtained from assigning entangled pairs
$|\Phi\rangle=\sum_{i=1}^D|ii\rangle$ to the edges of a lattice,
onto physical sites corresponding to the vertices. A natural
generalization of this concept to mixed states is to use
completely positive maps for the mapping from the virtual to the
physical level \cite{Ve04b}. Since every such map can be purified,
these mixed PEPS can be interpreted as pure PEPS with an
additional physical system which gets traced out in the end. To
become more specific let us consider a 2D square lattice. Then
every pure PEPS is characterized by assigning a 5'th order tensor
$A_{r,l,u,d}^i$ to each lattice site. Here the upper index
corresponds to the physical site and the lower `virtual' ones
(running from 1 to $D$) get contracted according to the lattice
structure. A mixed PEPS is then obtained by increasing the range
of $i$ from $d$ to $d d_E$ and finally tracing over these
additional environmental degrees of freedom, which can be thought
of as a second layer of the square lattice.

Let us now prove that all Gibbs states of Hamiltonians of
commuting finite range interactions are mixed PEPS. For simplicity
consider again a 2D square lattice. Starting point is to write the
un-normalized Gibbs state as $e^{-\beta H/2}\1 e^{-\beta H/2}$ and
to interpret the $\1$ as a partial trace over maximally entangled
states $|\Phi\otimes\Phi\rangle$ to which $e^{-\beta H/2}$ is
applied. In order to get an explicit form for the tensor $A$
assume that horizontally neighboring sites interact via $h_h$ and
vertical neighbors via $h_v$ and denote by
\be e^{-\beta h_v/2}=\sum_\alpha U_\alpha\otimes D_\alpha,\quad
e^{-\eta h_h/2}=\sum_\beta R_\beta\otimes
L_\beta\label{eq:BoltzmannDecomp} \ee
Schmidt decompositions in the Hilbert-Schmidt Hilbert space. That
is, the operators $U_\alpha, D_\alpha,R_\beta,L_\beta$ form four
sets of orthogonal operators, which by assumption commute with
each other but not necessarily among themselves (e.g.
$[U_1,U_2]\neq 0$). Using that the Gibbs state is up to
normalization a product of terms as in
Eq.(\ref{eq:BoltzmannDecomp}) leads then to its PEPS
representation with $D=d^2$ and
\be A_{r,l,u,d}^i = \big[L_rR_lU_dD_u\big]_{i_1,i_2}
\label{eq:AComGibbs}, \ee
where $i=(i_1,i_2)$ with $i_2$ corresponding to the environmental
degrees.

\subsection{Decay of correlations for Finitely correlated
states}

We consider now so called finitely correlated states (FCS)
\cite{FNW92}, which naturally appear in several lattice systems in
1D. They can be viewed as 1D PEPS (or matrix product states) where
all the local projectors are the same. Every FCS is most easily
characterized by a completely positive, trace preserving map (a
channel) $T:{\cal B(H}_1)\rightarrow {\cal B(H}_1\otimes {\cal
H}_2)$ with ${\cal H}_1,{\cal H}_2$ Hilbert spaces of dimension
$D, d$ respectively. Define further $\E(x)=\tr_2[T(x)]$ and assume
the generic condition that $\E$ has only one eigenvalue of
magnitude one, corresponding to a fixed point
$\varrho=\E(\varrho)$. The second largest eigenvalue, $\eta$, is
related to the standard correlation length through
$\xi=-1/\ln\eta$. With this notation, it is very simple to express
the states corresponding to regions $A$, $B$, and $AB$ (which are
required in order to determine the mutual information). We now
show that as $L$ gets larger, $\rho_{AB}$ approaches exponentially
fast $\rho_A\otimes \rho_B$.

The reduced density matrix $\rho_A$ of $N_A=R-L$ contiguous sites
is obtained as
\be \rho_A=\tr_1\Big[T^{N_A}(\varrho)\Big]. \label{eq:FCSrhoA} \ee
Similarly the joint reduced state of two regions $A$ and $B$ which
are separated by $L$ sites is given by
\be \rho_{AB}=\lim_{N_B\to\infty}\tr_1\Big[T^{N_B}\E^L T^{N_A}\E^L
T^{N_B}(\varrho)\Big]. \label{eq:FCSrhoAB} \ee
For sufficiently large $L$ write
\be \E^L(x) =\big(1-c \eta^L\big) \tr[x] \varrho +c \eta^L
\E'(x),\label{eq:EL} \ee
where $\E'$ is some channel and $c$ an $L$-independent constant.
Taken together Eqs.(\ref{eq:FCSrhoA}-\ref{eq:EL}) enable us to
bound the norm distance
\be ||\rho_{AB}-\rho_A\otimes\rho_B||_1 \leq 4c\eta^L
\label{eq:FCSnormbound} \ee
independent of $N_A,N_B$. That is, the two regions factorize
exponentially on a scale $\xi=-1/\ln\eta$ which can be regarded
the correlation length of the system. We cannot use this result
directly for the mutual information since the dimension of the
Hilbert space of system $B$ is infinite. However, we can proceed
by noting that each $T$ can be locally purified thereby increasing
the size of ${\cal H}_2$ by a factor of $d D^2$ (with $\E$
unchanged). Denoting the additional purifying systems by $A'$ and
$B'$ respectively, we obtain on the one hand $I(A:B)\leq
I(AA':BB')=S(\rho_{AA'}\otimes\rho_{BB'})-S(\rho_{AA'BB'})$. On
the other hand we can apply Fannes' inequality,
$|S(\rho)-S(\sigma)|\leq \Delta\log (\delta-1) +
H(\Delta,1-\Delta)$, where $\Delta=\frac12||\rho-\sigma||_1$ and
$\delta$ is the dimension of the supporting Hilbert space, to
$I(AA':BB')$. The advantage is that in this system we deal with
finite dimensional systems with $\delta=D^2$.

\newpage

\end{document}